\documentclass[]{piparticle-final}
\usepackage{graphicx}
\usepackage{amsmath}
\usepackage{cite}
\usepackage{color}

\usepackage{cite} % To compress citation ranges
\usepackage{epstopdf}

\begin{document}

\volume{8}               % To be inserted by Editor
\articlenumber{080001}   % To be inserted by Editor
\journalyear{2016}       % To be inserted by Editor
\editor{C. S. O'Hern}   % To be inserted by Editor
\reviewers{A. Baule, Queen Mary University of London, UK.}  % To be inserted by Editor
\received{2 November 2015}     % To be inserted by Editor
\accepted{{22} December 2015}   % To be inserted by Editor
\runningauthor{P. A. Gago \itshape{et al.}}  % To be inserted by Editor
\doi{080001}         % To be inserted by Editor

\title{Ergodic--nonergodic transition in tapped granular systems: \\The role of persistent contacts}

\author{Paula A. Gago,\cite{inst1,inst2,inst3}
        Diego Maza,\cite{inst4} 
        Luis A. Pugnaloni\cite{inst1,inst2}\thanks{E-mail: luis.pugnaloni@frlp.utn.edu.ar}}
        
\pipabstract{
Static granular packs have been studied in the last three decades in the frame of a modified equilibrium statistical mechanics that assumes ergodicity as a basic postulate. The canonical example on which this framework is tested consists in the series of static configurations visited by a granular column subjected to taps. By analyzing the response of a realistic model of grains, we demonstrate that volume and stress variables visit different regions of the phase space at low tap intensities in different realizations of the experiment. We show that the tap intensity beyond which sampling by tapping becomes ergodic coincides with the forcing necessary to break all particle--particle contacts during each tap. These results imply that the well-known ``reversible'' branch of tapped granular columns is only valid at relatively high tap intensities.
}

\maketitle

\blfootnote{
\begin{theaffiliation}{99}
   \institution{inst1} Dpto. Ingenier\'ia Mec\'anica, Facultad Regional La Plata, Universidad Tecnol\'ogica Nacional, Av. 60 Esq. 124, 1900 La Plata, Argentina.
   \institution{inst2} Consejo Nacional de Investigaciones Cient\'ificas y T\'ecnicas, Argentina.
   \institution{inst3} Current address: Department of Earth Science and Engineering, Imperial College London, South Kensington Campus, London SW7 2AZ, UK.  
   \institution{inst4} Departamento de F\'{\i}sica y Matem\'{a}tica Aplicada, Facultad de Ciencias, Universidad de Navarra, Navarra, Spain.
\end{theaffiliation}
}

\maketitle

\section{Introduction}

Granular matter is ubiquitous in nature. However, due to the complexity of the real particle--particle interactions, the standard approaches of continuum mechanics and thermodynamics are still limited in providing meaningful descriptions of the states in which these systems can be. Edwards and Oakeshot introduced a tentative approach inspired by the ideas of equilibrium statistical mechanics to formally describe the global properties of a static granular pack. Since the introduction of this theory ---where the entropy of the systems is governed by the spatial disorder of the grains \cite{edwards}---, a number of studies have used it to frame the interpretation of the results of specific experiments. The most relevant case is the so-called ``Chicago experiment'', where a column of grains was repeatedly tapped following an annealing-type protocol \cite{chicago,chicago2}. The main outcome of this experiment is that a stationary state can be reached, where the mean volume fraction, $\phi$, is a well defined 
function of the tap amplitude, $\Gamma$. Others have also obtained seemingly reproducible states without the need of annealing \cite{ribiere}. However, it has been shown recently, by simulation of frictionless grains, that these stationary states are not necessarily ergodic \cite{frenkel}. At low $\Gamma$, different members of an ensemble of steady-states prepared with a well defined protocol may sample a different region of the phase space, as the fluctuations of $\phi$ indicate.

In this paper, we demonstrate that not only the volume but also the force moment tensor, $\Sigma$, are sampled in a non-ergodic fashion and that ergodicity seems to be recovered if all particle--particle contacts are lost during each tap. This sets a clear limit to the range of driving forces able to generate a sequence of configurations for which the Edwards framework can be applied.

\section{Numerical protocol}
We simulated using the LAMMPS package \cite{plimpton2007} a quasi-two-dimensional cell containing $N=1000$ spherical particles of diameter $d$. The cell is $1.1\;d$ thick and $27.8\;d$ wide (the granular column is about 35 layers deep) to have a one to one representation of a previously introduced experimental device  \cite{pugnaloni2010,ardanza2014topological}. We use a model for soft frictional spheres described in Refs. \cite{brilliantov1996, silbert2001}. The normal component, $F_\mathrm{n}$, of the contact interaction is given by an elastic repulsive force proportional to the overlap of the interacting spheres and a dissipative term proportional to the normal component of the relative velocity. The tangential term, $F_\mathrm{t}$, implements an elastic shear force and a damping force. The shear force takes into account the cumulated tangential displacement between the particles while they remain in contact. Whenever $F_\mathrm{t} > \mu F_\mathrm{n}$ ($\mu$ is the 
friction coefficient), this lower dynamic friction force 
is used. In this work we use the same interaction parameters as in Ref. \cite{pugnaloni2008,pugnaloni2011}. The wall--particle interaction is defined with the same parameters as the particle--particle force. Tapping is simulated by imposing an external vertical motion to the cell. This pulse is a single sinusoidal cycle $A \sin(\omega t) $. We fix $\omega=2\pi\times33$ Hz and use the tap amplitude $A$ as control parameter. The tap intensity is characterized by $\Gamma=A\omega^2/g$. The mechanical equilibrium after each tap is deemed achieved if the kinetic energy of each particle has fallen (in average) below $10^{-6} mgd$. Where $m$ is the mass of one particle and $g$ the acceleration of gravity.

We study 20 independent realizations of a decreasing ramp of the tap amplitude. We initially fill the cell by placing the spheres at random positions before letting them deposit under the action of gravity. In each realization, we decrease $\Gamma$ in small steps, from $\Gamma=20.0$ down to $\Gamma=0.8$, and apply $200$ taps for each $\Gamma$. Note that for $\Gamma<1.0$, the column of grains does not detach from the base during a tap. The $200$ taps at each value of $\Gamma$ are enough to reach a steady-state. We do not observe any drift of the mean values of $\phi$ or $\Sigma$ after the initial $100$ taps, which we will discard later in our analysis. Finally,  we also study a cyclic annealing protocol: starting from the final configuration at $\Gamma=0.8$ for each of the former realizations; the tap amplitude is cyclically increased and decreased every 200 taps in the range $0.8<\Gamma<5.5$ in order to compare the steady-states reached with an alternative method.

\section{Data analysis}
To measure the packing fraction we use the 2D Voronoi tessellation (implemented in \cite{voro++}) of the $x$--$z$ plane projection of the particle positions, disregarding the third coordinate on the thin direction of the cell. Then, we associate to each particle a ``local volume'' fraction by dividing the particle area by the corresponding Voronoi area. In order to avoid boundary effects, we disregard particles closer than $2d$ to the lateral walls. Following the recommendations in Ref. \cite{gago2015}, we analyze horizontal slices of the granular column $15\;d$ thick measured at approximately the same depth with respect to the free surface in order to retrieve unbiased results for the force moment tensor due to the uneven free surface. Averaging over the $N$ particles contained in the slice of interest, we obtain the volume fraction of each static configuration. To obtain the steady-state $\phi$ corresponding to a given $\Gamma$, we averaged this quantity over the last 
$100$ configurations obtained for each tap intensity. We 
also obtain the force moment tensor $\sigma_{i}^{\alpha\beta}= \sum \limits_{c} r_c^{\alpha} f_c^{\beta}$ of each particle in the slice of interest. Here, the sum runs over all the contacts $c$ of the particle, $\overrightarrow{r}_c$ is the vector from the center of the grain to the contact $c$ and $\overrightarrow{f}_c$ is the corresponding contact force. We apply the same averaging protocol used for $\phi$ to obtain the force moment tensor for the configuration and $\Sigma$ for the steady-state of a given realization and $\Gamma$.

 \begin{figure}[t!]
\includegraphics[width=1\columnwidth,trim= 0cm 0cm 0 0cm,clip]{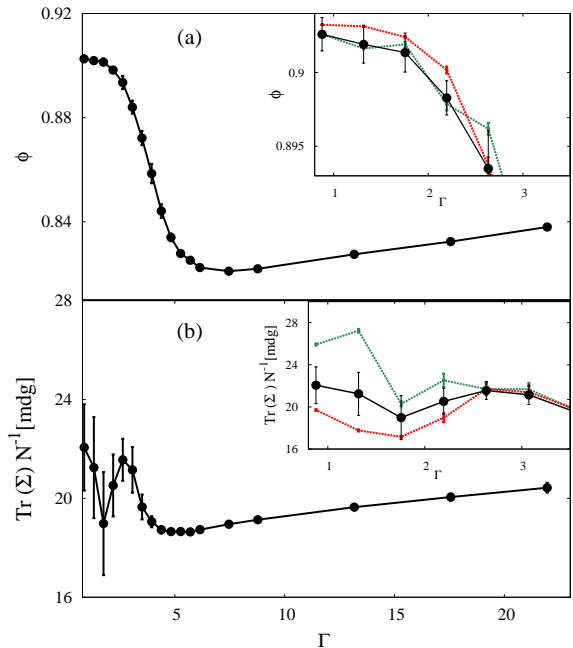}
\caption {Ensemble average of the steady-state packing fraction $\phi$ (a) and trace of the force moment tensor $\rm{Tr}(\Sigma)$ (b) as a function of $\Gamma$. The error bars correspond to the standard deviation over the $20$ averaged realizations. The insets show the same results and two of the $20$ independent realizations (dashed lines) in the low-tap-intensity region. The error bars on the single realization data correspond to the estimated standard error of the mean.}
\label{Fig1}
\end{figure}

\begin{figure}[t]
\includegraphics[width=1\columnwidth,trim= 0cm 0cm 0 0cm,clip]{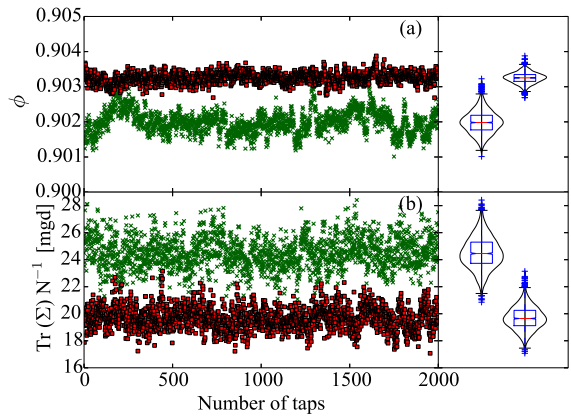}
\caption {$\phi$ (a) and $\rm{Tr}(\Sigma)$ (b) as a function of the tap number for two of the $20$ independent realizations at $\Gamma=0.8$. The notched boxes and ``violin'' diagrams shown suggest that both realizations can be hardly considered as representing the same steady-state.}
\label{Fig2}
\end{figure}

\section{Results}
In Fig.~\ref{Fig1}(a) the ensemble  average of the steady-state $\phi$ (i.e., averaged over the 20 independent realizations) is displayed as a function of $\Gamma$. The error bars correspond to the standard deviation over the 20 realizations. As observed by a number of authors, the curve seems to be very well defined with independent realizations falling within a very narrow range of $\phi$ values for any given $\Gamma$. In the past, this led to the conclusion that this was a truly reversible process, where lowering or raising $\Gamma$ would lead to the same steady-state $\phi$. In the inset of Fig.~\ref{Fig1}(a) two of the independent realizations are shown for the low-tap-intensity region. From this picture, it is clear that steady-states corresponding to a given $\Gamma$ can differ from one realization to another. Notice that in the inset the error bars for the two isolated realizations correspond to the standard error of the mean (SEM), which gives an estimate of the uncertainty of the mean 
value reported rather than the size of the $\phi$ fluctuations. For these two realizations, although the mean $\phi$ seems to agree within the estimated error for intensities $\Gamma>1.5$, it is clear that they are different for low $\Gamma$. This is consistent with the findings of Paillusson and Frenkel \cite{frenkel} for frictionless spheres under event-driven simulations. However, in our simulations we are able to extract the stress state of the system as well as the history of the contacts. These reveal valuable information, as we discuss below.

In Fig.~\ref{Fig1}(b), we show the trace $\rm{Tr}(\Sigma)$ of $\Sigma$ averaged over all 20 realizations as a function of $\Gamma$. As before, the error bars indicate the standard deviation over the 20 realizations. As we can see, the variability of the mean stress is significantly large at low $\Gamma$. This is not due to the large fluctuations during a given series of taps but to the variations observed from one realization of the protocol to the other. Indeed, the inset in Fig.~\ref{Fig1}(b) shows that, for low $\Gamma$, the mean values of $\rm{Tr}(\Sigma)$ for two of the 20 realizations have a relatively small SEM (i.e., the fluctuations in each realization are small). However, realizations differ from each other. The difference between results corresponding to different realizations become much more evident here than in the case of the $\phi$--$\Gamma$ plot. We suggest that the stress tensor may be more sensitive and then more suitable to sense if ergodicity is fulfilled in experimental data. Overall, as 
$\Gamma$ is decreased, different realizations explore non-overlapping ranges of volume/stress. Therefore, temporal averages (on a single time series) do not match with ensemble averages (over the realizations).

\begin{figure}[t!]
\includegraphics[width=1\columnwidth,trim=0mm 0mm 0mm 0mm, clip]{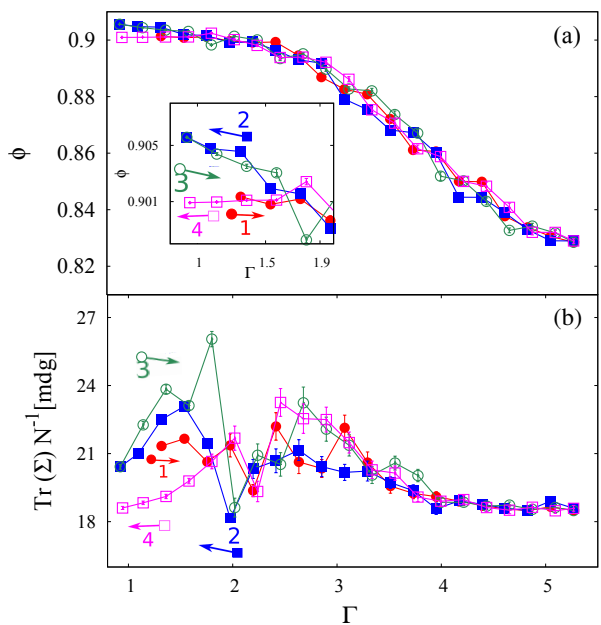}
\caption {Steady-state $\phi$ (a) and $\rm{Tr}(\Sigma)$ (b) as a function of $\Gamma$ corresponding to a full annealing protocol. Starting from the filled circles (red) increasing ramp, and following by filled squares (blue), open circles (green) and open squares (magenta). The error bars correspond to the estimated SEM.}
\label{Fig3}
\end{figure}

Since the number of taps we have explored for each $\Gamma$ may be small to assume that the steady-state has been properly sampled, we carried out $2000$ additional taps at $\Gamma=0.8$ for each realization. Since some of the signals are not normally distributed, we confirmed the stationarity of these states by using a non-parametric test at a level of significance of $5 \%$ \cite{fractal,R}. Two of the normally distributed realizations are displayed in Fig.~\ref{Fig2} as a function of the tap number. Comparing the corresponding notched boxes and ``violin'' diagrams of both signals, it is clear that the states do not match. Hence, the \emph{reversible branch} found in Ref. \cite{chicago2} is not so at low $\Gamma$ in our case since truly stationary states with distinguishable $\phi$ and $\rm{Tr}(\Sigma)$ may be obtained on different realizations of the annealing protocol. Of course, this is harder to detect in $\phi$ since the dispersion between realizations is much smaller compared with the 
range of $\phi$ values obtained at different $\Gamma$.

The former results also confirm the hypothesis that non-ergodicity is present in a typical tapping protocol beyond the special case reported in Ref. \cite{frenkel}, where the steady-states obtained did not follow any annealing-type protocol. Hence, we observe this non-ergodic behavior even after annealing the system from high tapping strengths. To stress this point and assess if the speed of the annealing may prevent the system from reaching a unique steady-state on each realization, we apply a slower cyclic annealing protocol (similar to the one introduced in \cite{chicago}) to each of the 20 final states at low $\Gamma$ in order to reproduce the ``reversible branch''. In Fig.~\ref{Fig3} we display a sequence of two successive up and down ramps applied to one of the 20 initial realizations using $\Gamma$-steps about one half of those used in Fig. \ref{Fig1}. Although in the scale used for $\phi$ the steady-state packing fraction seems reversible, a close inspection shows that the states have distinguishable 
$\phi$ at low $\Gamma$ [see Fig. \ref{Fig3}(a) and the corresponding inset]. This is much more evident when the stress is analyzed [see Fig.~\ref{Fig3}(b)]. %We mention in pass the interesting case of the two blue ramps (one corresponding to increasing and the other to decreasing $\Gamma$) which have the peculiarity that they seem to be sampling the same steady-state as shown by the mean $\phi$ obtained at low $\Gamma$. However, the corresponding values for $\rm{Tr}(\Sigma)$ are clearly distinguishable, indicating that the steady-states for each $\Gamma$ on these two ramps are in fact different.

\begin{figure}[t!]
\includegraphics[width=1\columnwidth,trim= 0cm 0cm 0 0cm,clip]{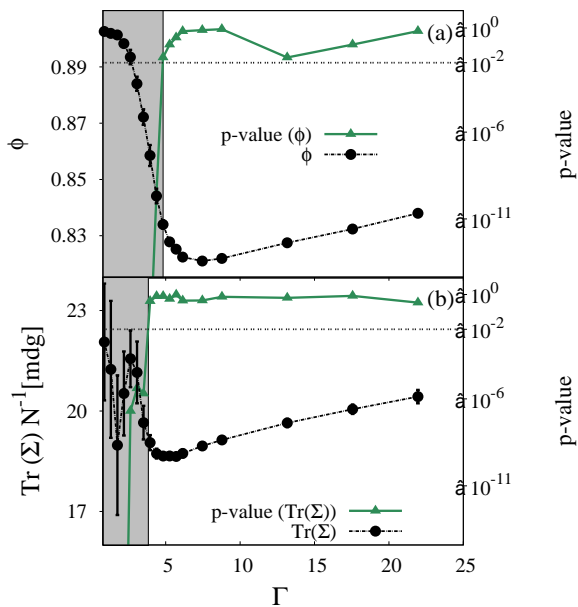}
\caption {P-value (up-triangles, right axis) as a function of $\Gamma$ for the Kruskal--Wallis \cite{kruskal} one-way analysis of variance for $\phi$ (a) and $\rm{Tr}(\Sigma)$ (b). The horizontal dotted line corresponds to the significance level used (1\%). The black circles correspond to the $\phi$ and $\rm{Tr}(\Sigma)$ data from Fig.~\ref{Fig1}.}
\label{Fig4}
\end{figure}

In order to set a criterion to decide if the steady-states are not ergodic for a given $\Gamma$, we show in Fig. \ref{Fig4} the $p$-values for the Kruskal--Wallis \cite{kruskal} one-way analysis of variance performed on the 20 realizations at each $\Gamma$. This simple non-parametric test allows for the rejection of the null hypothesis that all 20 data series are drawn from a unique distribution (which does not need to be normal), hence that they correspond to a unique steady-state. If the $p$-value is significant (in our case $p>0.01$), then we cannot rule out the possibility that the 20 series come from the same steady-state. As we can see from Fig. \ref{Fig4}(a), the test run on the data for $\phi$ indicates that for $\Gamma<5.0$, the null hypothesis must be rejected and therefore there exist at least two out of the 20 steady-states that are not the same. However, for higher $\Gamma$, the test is significant and then the 20 realizations may correspond to the same steady-state. Interestingly, when the test 
is 
run on $\Sigma$ [see Fig. \ref{Fig4}(b)], the steady-state seems to be unique for all 20 realizations if $\Gamma>3.75$. Although differences between realizations are simpler to detect on visual inspection of $\rm{Tr}(\Sigma)$, it is actually $\phi$ that sets a higher threshold for the $\Gamma$ values needed to ensure an ergodic steady-state (i.e., $\Gamma>5.0$).

\begin{figure}[ht!]
\includegraphics[width=1\columnwidth,trim= 0cm 0cm 0 0cm,clip]{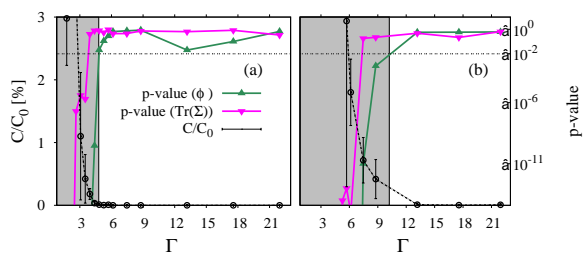}
\caption {(a) Percentage $C/C_0 \times 100$ of persistent contacts (black circles) as a function of $\Gamma$, averaged over $5$ taps on $6$ independent realizations. The error bars correspond to the standard deviation over the $6$ realizations. Up-triangles (green) correspond to the p-values in Fig.~\ref{Fig4} for $\phi$ and down-triangles (magenta) to $\rm{Tr}(\Sigma)$. (b) Same as (a) for frictionless grains.}
\label{Fig5}
\end{figure}

The previous results indicate that the steady-states sampled at low tap intensities do not only depend on $\Gamma$ but on the particular history of each realization. Notice that this goes beyond the history dependent out-of-equilibrium trajectories already reported in tapped systems \cite{josserand} since here we are focusing on the steady-states. One may hypothesize that the constraints imposed by the contacts is one of the reasons for the non-ergodic behavior at low tap. If a contact persists from one tap to the next, the contact force after coming back to rest will depend on the history of all contacts of that particular grain. In order to test this idea, we analyze the evolution of all contacts during each tap and identify those that persist (i.e., contacts that did not break at any time during the pulse of energy). Figure ~\ref{Fig5} shows the average ratio $C/C_0$ of persistent contacts, $C$, to the total number of contacts, $C_0$, as a function of $\Gamma$. For this calculation, each contact was 
tracked during the final $5$ taps for each $\Gamma$ on $6$ of the independent realizations and only grains that fall within the layer of interest, as discussed above, where included in the analysis. The percentage of persistent contacts is very small but non-zero up to $\Gamma \approx 5.0$. As it is expected, when $\Gamma$ is increased sufficiently all the contacts are broken and new ones are made during each tap (resulting in $C/C_0=0$). This transition coincides with the value of $\Gamma$ where the realizations seem to sample the same steady-state (see the $p$-values included in  Fig.~\ref{Fig5}). Therefore, when small taps are applied, the aging of some of the contacts seem to lead the system to sample different regions of the phase space during independent realizations. However, if all contacts are made anew at each tap, the sampling becomes compatible with the idea of ergodicity introduced in Fig.~\ref{Fig4}. In order to generalize this result, we also simulated 
frictionless grains. Interestingly, the same conclusion drawn for frictional grains is true for frictionless ones: different realizations seem to sample the same steady-state only if all contacts are made anew upon each tap [see Fig.~\ref{Fig5} (b)].

\section{Conclusions}
Our analysis of the steady-states of tapped granular systems indicate that these states are history-dependent for tap intensities below a certain threshold. This is in contradiction with the general assumption that macroscopic time averages ---such as the volume fraction--- can be recovered when the amplitude of the perturbation applied to the system is tuned back and forth. The differences between independent realizations become particularly noticeable in the stress distribution. These findings show that the postulates of the equilibrium statistical thermodynamics may not be always fulfilled to describe the steady state of static granular systems (see also Ref. \cite{irastorza} for a discussion on the Boltzmann distribution failure for an analytically solvable model). Focusing on tap intensities that warrant that all contacts are made anew after each tap may allow exploring the available phase space in agreement with the ergodic hypothesis. However, gentle perturbations deserve an approach that includes memory effects to suitably describe the states. 
In that sense, non-equilibrium thermodynamic approaches may be a suitable alternative \cite{lebon}.  Further research on such alternative formalisms, the effect of other types of forcing mechanisms (e.g., shear), and possible extensions to other complex systems (e.g., active matter) become necessary.

\begin{acknowledgements}
This work has been partially supported by projects PICT-2012-2155 ANPCyT (Argentina), FIS2011-26675 MINECO (Spain) and PIUNA (Universidad de Navarra).
\end{acknowledgements}

\end{document}